\documentstyle[12pt]{article}

\renewcommand{\thesection}{\arabic{section}.}
\renewcommand{\theequation}{\thesection\arabic{equation}}
\renewcommand{\thesubsection}{}

\def\subsect{\setcounter{equation}{0}\subsection}
\def\sect{\setcounter{equation}{0}\section}

\def\appendix{\section*{{\hfill{Appendix}\hfill}\markright{Appendix}}
\setcounter{subsection}{0}
\renewcommand{\thesubsection}{\Alph{subsection}.}
\renewcommand{\theequation}{\thesubsection\arabic{equation}}}

\title{
{\large\bf 
LOW TEMPERATURE DOMINANCE OF PION-LIKE EXCITATIONS IN THE
MASSIVE GROSS-NEVEU MODEL AT ORDER 1/N\thanks{Partially 
supported by the Swiss National Foundation}}  
\vspace{1cm}}
\author{\large
A.Barducci, R.Casalbuoni, M.Modugno, G.Pettini\\
{\small Dipartimento di Fisica, Univ. di Firenze and
I.N.F.N., Sezione di Firenze }\\[0.5cm]
R.Gatto\\
{\small D\'epartement de Physique Th\'eorique, Univ. de Gen\`eve}
\vspace{1cm}}
\date{DFF 255-7-96}

\begin{document}

\maketitle
%--------------------------------------------------
%- levare (mettere) i commenti qui sotto se si usa 
%- (non si usa) il logo dell'universita' di Ginevra
%---------------------------------------------------
% \thispagestyle{empty}
% \newpage 
%---------------------------------------------
%---------------------------------------------

\vspace{1cm}
\begin{abstract}
\indent\noindent

We perform a $1/N$-expansion of the partition function 
of the massive Gross-Neveu model in 1+1 dimensions.
The procedure allows for the inclusion  of the 
contribution of scalar and 
pseudoscalar composites (of order $1/N$)
to the equation of state.
The naive expectation that the bosonic fluctuations
correct significantly the mean field approximation
at low temperatures is confirmed by our calculations. 
Actually the relevant degrees of freedom 
of hadronic matter at low temperatures are found to be 
pion-like excitations, rather than the fundamental 
constituents.

\end{abstract}
\newpage

\section{Introduction}
\indent\noindent

The equilibrium thermodynamics of strong-interacting matter 
has been studied, up to now, by quite
different approaches.
Apart from lattice simulations, most of the other studies
rely on phenomenological models, which try to capture some relevant
physical features of the theory \cite{bard,pisarski,brown}.
The absence of a common starting point in the literature is due 
to the non perturbative character of low energy QCD,
and to the vicinity of the expected phase transition, 
at temperatures of the order of $\Lambda_{QCD}$.
Thus, in employing analytical methods to study the
transition, one is faced by this major difficulty of QCD, 
that of passing from the hadronic regime to that of
quarks and gluons, based on the same fundamental 
lagrangian.

For very low temperatures, important results 
have been obtained on the assumption that the 
thermodynamics of hadronic matter in this regime 
is that of a gas of weakly interacting pions \cite{leutw},
since pions are the lowest mass 
excitations and are thus expected to dominate the partition function.
On the other hand, at very high temperatures, 
we expect a gas of almost free quarks and gluons.
Results obtained in these limits,
which are both far from the temperatures
of deconfinement and chiral symmetry restoration,
cannot help to explore the nature of the transition itself.

For such an exploration much of the effort done so far is based on the
mean field approximation to the effective potential at
finite temperature for four-fermion models or other
low-energy effective models of QCD.
Unfortunately, the mean field approximation neglects
fluctuations, and it misses completely,
in such a way, the role of bosonic contributions
to the thermodynamical potential. These however
should be, as already said, the relevant ones at very low
temperatures.

Recently there have been attempts to recover some
information missing in the mean field approximation,
by means of the $1/N$ expansion, where $N$ labels
the number of fermion species \cite{kleva,nexp}. 
In fact the role
of scalar and pseudoscalar fluctuations
appears at the first order of the expansion and 
allows to test their relevance in various regions of temperature.
For the moment being these attempts have been
made for   the Nambu-Jona Lasinio (NJL) model \cite{njl}.

Here we present an application, based on a general scheme
which allows for the $1/N$ expansion of the effective action in
four-fermion models \cite{unosuenne}, to the Gross-Neveu 
model in 1+1 dimensions \cite{gn}. 
We include in the Lagrangian a current fermion mass to avoid infrared 
divergencies, and include the one pion-like 
composite states and the scalar composite.
To verify whether their contribution to the free energy is 
dominant at low temperatures, we have
to evaluate the effective potential at least at order $1/N$. 
In the imaginary time formalism,
we have to sum up over discrete Matsubara frequencies
twice: the first time for fermions, the second one
for bosons.

The effective potential  
at this order of the expansion appears as a sum of various terms. 
The first is the zeroth order, mean field term. It has a zero temperature
part and a finite temperature part, obtained by summing over
fermionic energies.

A second term is a pure temperature term,
obtained by summing over both fermion and boson energies.
It contains a Bose distribution function and it is 
expected to dominate in the expression for the pressure at low temperatures. 
In the following we show this to be indeed
the case, and that in particular the largest contribution
comes from the integration over almost on shell external pion momenta
(close to the pion propagator pole).

The last term is a mixed one, a zero temperature term with 
respect to the sum over Bose frequencies, 
depending on temperature through  the bosonic
propagator, obtained by summing over Fermi frequencies. 

This part of the effective potential presents non trivial
aspects which are interesting as far as the
model itself is concerned, but its  detailed study
would go beyond the scope of the present 
work. Actually we are primarily interested
in showing how, by applying a general formula for  
four fermion theories \cite{unosuenne}, one 
can evaluate the pressure at order $1/N$, and exhibit 
the role of pions at low temperatures.
Since the mixed term is subleading for very low temperatures
and small quark masses (see sect.5), 
it does not appear in our final results.

In the following paragraph we give a review of the $1/N$ expansion
of the partition function, for the case of the Gross-Neveu model
in 1+1 dimensions. 
In sect.3 and sect.4 we extend the formalism to finite temperature,
 and calculate the pressure at order $1/N$.
In sect.5 we show the temperature behaviour of the pressure 
for the pure
bosonic term discussed above and  compare it to the mean field term.
The  results are summarized in the conclusions.

Finally we summarize some useful calculations in the appendices, namely
the analytic properties of the inverse bosonic propagator (App.A),
and the method employed for summing over the bosonic frequencies (App.B).

\sect{$1/N$ expansion}
\indent\noindent

We start from the lagrangian of the massive Gross-Neveu model 
in $1+1$ dimensions~\cite{gn}
\begin{equation}
{\cal L}=\bar{\psi}\left( i \hat{\partial}-M\right)\psi 
+ {g^2 \over2} \left[(\bar{\psi}\psi)^2 -
(\bar{\psi}\gamma_5\psi)^2\right]
\end{equation}
where $\psi$ is a multiplet of $N$ degenerate fermion fields, and
$M$ is  a bare fermion mass which explicitly breaks the chiral invariance.

The $1/N$ expansion can be carried out as follows \cite{unosuenne}.
According to standard methods  the generating functional can 
be rewritten by introducing a scalar field $\sigma$ and a 
pseudoscalar field $\pi$ \cite{gn}. At vanishing sources, we have
\begin{equation}
Z =\int{\cal D}(\bar{\psi}\psi){\cal D}(\sigma){\cal D}(\pi)~
e^{\displaystyle i\int {d^2 x}
\left[\bar{\psi}(i\hat{\partial}- M + g\sigma
+i g\gamma_5\pi)\psi -
{1 \over 2}(\sigma^2+\pi^2)\right]}
\end{equation}

By carrying out the integral over the fermion fields, we obtain an effective
action for scalar and pseudoscalar fields
\begin{equation}
Z =\int{\cal D}(\sigma){\cal D}(\pi)~
e^{\displaystyle i S_{B}}\equiv ~ e^{\displaystyle i W}
\label{zeta}
\end{equation}

Now we shift the $\sigma$-field, 
$g\sigma\rightarrow g\sigma+M$, and write
\begin{equation}
S_{B}=\int {d^2 x}\left[-{1\over 2}(\sigma^2+\pi^2)
 (1+\delta Z) - {M\sigma\over g}\right]-i\log {\rm det}
(i\hat{\partial} +g\sigma+i g\gamma_5\pi)
\end{equation}
where $\delta Z$ is a counterterm which 
renormalizes  the fermion loop \cite{gn}.

Finally we redefine the parameters of the model as follows
\begin{equation}
\lambda\equiv Ng^2 \; ;\quad
\alpha\equiv {M \over Ng^2} \; ;\quad
\phi\equiv(g\sigma,g\pi)\; ;\quad
a\equiv (1,i \gamma_5)
\end{equation}

Thus, the lagrangian for the $\phi$-field is given by
\begin{eqnarray}
S_{B} =\int {d^2 x}\left[ -{N\over 2\lambda}\phi^2 (1+\delta Z)
- N\alpha\phi_{1}\right]
-iN{\rm Tr}\log(i\hat{\partial} +\phi\cdot a)
\equiv\int {d^2 x}{\cal L}_B(\phi)\label{boson}
\end{eqnarray}
where we have used the identity
{\sl log  det =  tr  log},
and carried out  the trace over fermion indices.

Up to now we have simply integrated over the fermion fields, 
passing from a description in terms of elementary fermionic 
constituents to a description in terms of bosonic composites.
At this point we are ready to expand the generating 
functional $Z$ in eq.(\ref{zeta}) in series of $1/N$.
For this we follow the method of ref.\cite{jackiw},
that we briefly review (notice that we are now performing
an $1/N$ expansion and not the  formal $\hbar$-expansion
of the original reference).

Let us couple an external constant source $J$ to the field $\phi(x)$,
and expand $\phi(x)$ around its vacuum expectation value $\bar{\varphi}$
\begin{eqnarray}
\phi(x)&=&\phi'(x)+\bar{\varphi}\label{phiprimo}\\
\bar{\varphi}[J]&=&\langle0^+|\phi(x)|0^- \rangle_J = 
{\delta W\over\delta J}
\label{barphi}
\end{eqnarray}

Since $J$ is constant, from the Lorentz invariance of the vacuum, it 
follows that $\bar{\varphi}$ is constant too. By supposing $\bar{\varphi}[J]$ 
to be invertible, we can define the effective action 
$\Gamma[\bar{\varphi}]$ 
as the Legendre transform of $W[J]$
\begin{equation}
\Gamma[\bar{\varphi}] = W[J[\bar{\varphi}]] - 
\int {d^2 x}J[\bar{\varphi}]\bar{\varphi} \equiv -{\cal V}
(\bar{\varphi})\int {d^2 x}\label{gamma}
\end{equation}
From the definitions it follows that, at $J=0$, 
$\bar{\varphi}$ must satisfy the 
stationary equation
\begin{equation}
{\delta {\cal V}({\varphi})\over
\delta{\varphi}}\Bigg|_{{\varphi}=\bar{\varphi}}=0
\label{gapw}
\end{equation}

At this point $W[J]$ can be separated into two terms
\begin{equation}
W[J]\equiv W_{0}[J]+W_{1}[J]
\end{equation}
with  $W_0[J]$ given by
\begin{equation}
W_{0}[J]=\int {d^2 x}\Big({\cal L}_{B}(\bar{\varphi})+J\bar{\varphi}\Big)
\label{w0}
\end{equation}
and $W_1[J]$ solution of the integral equation \cite{jackiw}
\begin{equation}
W_{1}[J]\equiv -i\log\int{\cal D}(\phi')
\exp\left[i\int{d^2 x}\left({\cal L}_{B}^{(2)}(\phi',\bar{\varphi})-\phi'
{\delta W_1\over\delta\bar{\varphi}}\right)\right]\label{wuno}
\end{equation}
The lagrangian ${\cal L}_{B}^{(2)}(\phi',\bar{\varphi})$ is 
${\cal L}_B(\phi'+\bar{\varphi})$ minus the constant and
linear terms in $\phi'$, and can be obtained from
eq.(\ref{boson}). We have 
\begin{equation}
S_{B}=\int {d^2 x}\left[ -{N\over 2\lambda}(\phi'+\bar{\varphi})^2 
(1+\delta Z)
- N\alpha(\phi_{1}'+\bar{\varphi}_{1})\right]
-iN{\rm Tr}\log(i\hat{\partial} +\bar{\varphi}\!\cdot\! a+\phi'\!\cdot\! a)
\label{sbos}
\end{equation}

Furthermore, by defining 
\begin{equation}
(i\hat{\partial}+\bar{\varphi}\cdot a)G(x_1-x_2)=\delta^2 (x_1-x_2)
\end{equation} 
it follows
\begin{equation}
{\rm Tr}\log(i\hat{\partial} +\bar{\varphi}\cdot a+\phi'\cdot a)=
{\rm Tr}\log(i\hat{\partial} +\bar{\varphi}\cdot a)
+{\rm Tr}\log\left(1+G\phi\cdot a\right)
\end{equation}
Therefore, the standard mean field fermionic term in eq.(\ref{w0}), 
at $J=0$, is
\begin{eqnarray}
W_{0}&=&\int {d^2 x}\left[ -{N\over 2\lambda}\bar{\varphi}^2 
(1+\delta Z)
- N\alpha\bar{\varphi}_{1}\right]
-iN{\rm Tr}\log(i\hat{\partial} +\bar{\varphi}\cdot a)
\nonumber\\
&=&\int {d^2 x}\left[ -{N\over 2\lambda}\bar{\varphi}^2 
(1+\delta Z)
- N\alpha\bar{\varphi}_{1}+
iN\int{d^2 p\over (2\pi)^2}\log(\bar{\varphi}^2-p^{2})\right]
\end{eqnarray}
This is the only term which survives in the $N\rightarrow+\infty$
limit, since $W_1$ is of order $1/N$ with respect to $W_0$.

The evaluation of $W_{1}$ requires to expand $S_B$ in eq.(\ref{sbos}) 
in powers of $\phi'$, starting from the quadratic terms. 
By rescaling the fields, $\phi'\rightarrow\phi'/\sqrt{N}$, and by 
 ordering the terms in the exponent of eq.(\ref{wuno}) in powers of $1/N$,
it turns out that the leading term is just the quadratic one.

Thus, by keeping only the leading term of the $1/N$ expansion, the
field-dependent part of $W_{1}$ can be obtained by
\begin{eqnarray}
e^{\displaystyle {iW_{1}}}&\equiv&  \int{\cal D}(\phi')
\exp\left\{-{i\over 2\lambda}
\int{d^2 x}\phi'^2 (1+\delta Z)\right.\nonumber\\
&&\qquad \left.-{1\over 2}
\int{d^2 x_1}{d^2 x_2}{\rm Tr}
\Big[G(x_1 - x_2)\phi'(x_1)\cdot a G(x_2-x_1)
\phi'(x_2)\cdot a\Big]\right\}
\end{eqnarray}

By going over to the Fourier space, 
we can diagonalize the exponent in the previous 
equation, thus obtaining a simple gaussian integral. 

We finally have
\begin{equation}
W_1
= \sum_{j=\sigma,\pi}{i\over2}\int{d^2 x}\int{d^2 p\over(2\pi)^2}
\log\left[{i\over2\lambda}D^{-1}_{0,j}(p)\right]
\label{w1}
\end{equation}
where $iD^{-1}_{0,j}(p)$ is the zero temperature inverse bosonic
propagator, whose explicit form is given in Appendix $A$.

At this point it is straightforward to verify that, 
apart from the linear term in $\bar{\varphi}_1$,
both $W_{0}$ and $W_{1}$ depend on the chirally invariant combination
${\bar{\varphi}}^2=g^2({\bar{\sigma}}^2+{\bar{\pi}}^2)$, which
is the solution of eq.(\ref{gapw}).
From this condition it follows that the pseudoscalar component
vanishes, i.e. 
\begin{equation}
\bar{\varphi}=(g\bar{\sigma},0)
\end{equation}

\sect{Finite temperature formalism}
\indent\noindent

The finite temperature extension is
straightforward. For imaginary times we have
\begin{equation}
\int{d^2 x}\longrightarrow-i\beta\int{d x}\equiv-i\beta V\; ;\qquad
\int {d^2 p\over (2\pi)^2}
\longrightarrow{i\over\beta}\sum_{n=-\infty}^{+\infty}
\int{d p\over 2\pi}
\end{equation}
while the finite temperature effective potential, 
obtained as the generalization
to finite temperatures of eq.(\ref{gamma}), 
is just the free energy density ${\cal F}$
\begin{equation}
  {\cal V}\longrightarrow -i{W\over \beta V}
  =-{\log Z\over \beta V}\equiv{\cal F}
\end{equation}

According to the $1/N$ expansion discussed in the previous paragraph, 
we can write
\begin{equation}
{\cal F} = {\cal F}^F + \sum_{j=\sigma,\pi}{\cal F}^B_j
\label{state}
\end{equation}
where ${\cal F}^F$ is the purely fermionic standard mean field
term coming from the one-loop calculation ($\omega_n=(2n+1)\pi/\beta$)
\begin{equation}
{\cal F}^F = {N\over 2\lambda}\bar{\varphi}^2(1+\delta Z)
 + N\alpha \bar{\varphi}
-{N\over\beta}\sum_n
\int{d p\over2\pi}
\log\left(\bar{\varphi}^2+p^2+\omega^2_n\right)
\label{mean}
\end{equation}
and ${\cal F}^B$ is the contribution of bosonic fluctuations
 ($\nu_n=2n\pi /\beta$)
\begin{equation}
{\cal F}^B_j = {1\over2\beta}\sum_n\int{d p\over 2\pi}
\log\left[{i \over2\lambda}D^{-1}_j (i\nu_n,p)\right]
\label{free}
\end{equation}

The expression of the finite temperature inverse bosonic 
propagator $iD^{-1}$ 
and its analytic properties  are given 
in Appendix $A$.

Expression (\ref{free}) and (\ref{mean}) put in eq.(\ref{state}) 
give the equation of state, when evaluated
at the solution of the gap equation (\ref{gapw})
at finite temperature
\begin{equation}
{\partial {\cal F}(\varphi,T)\over\partial \varphi}
\Bigg|_{\varphi=\bar{\varphi}(T)}=0
\label{gap}
\end{equation}
which determines the evolution
of the fermion condensate with temperature.

By following a standard renormalization procedure \cite{gn} and
evaluating the sums in the last term of eq.(\ref{mean}) \cite{dolan},
 the fermionic term
${\cal F}^F$ can be written as
\begin{equation}
{\cal F}^F(\varphi,T) = N\left[ {\varphi^2 \over 4\pi}
\left(\log{\varphi^2 \over m_0^2}
-1\right)+\alpha\varphi- 
 {2\over\pi\beta}
\int_0^{+\infty}\!\!\!\!{d p}\log\left(1+e^{\displaystyle 
-\beta\sqrt{p^2+\varphi^2} }
\right) \right]
\end{equation}
where $m_0$ is the dynamical fermion mass generated at $\alpha=0$, $T=0$.

The sums that appear in the bosonic part of the free energy, 
eq.(\ref{free}),
 can be transformed into integrals by standard
methods \cite{land}. 
In  this procedure 
 we must pay attention to the fact that the  finite temperature
inverse bosonic propagator $iD^{-1}(\omega,p)$ has a cut 
across the origin in the complex $\omega$-plane (for $|\omega|<p$),
besides the cut extending from the continuum threshold to $+\infty$
(that is present even at $T=0$) \cite{kleva,cut}. 
A detailed derivation is shown in Appendix $B$,
from which it follows that the bosonic term can be cast in the form
($n_{B}(\omega)=({\rm e}^{\displaystyle\beta\omega}-1)^{-1}$)
\begin{eqnarray}
{\cal F}^B_j&=&
{1\over2\pi}\int_{-\infty}^{+\infty}
{d p\over2\pi}\int_{0}^{+\infty}\!\!\!\!\!\! 
d\omega 
\log\left[i D_j^{-1}(i\omega)\right]
\nonumber\\
&&+\lim_{\epsilon\rightarrow0}{1\over2\pi i}
\int_{-\infty}^{+\infty}{d p\over 2\pi}
\int_0^{+\infty}\!\!\!\!\!\!{d\omega}~
n_{B}(\omega)\log\left[{i D_j^{-1}(\omega+i\epsilon,p)\over i 
D_j^{-1}(\omega-i\epsilon,p)}\right]
\end{eqnarray}
apart from infinities independent of $\beta$ and $\varphi$.

Furthermore, since the function $g(z)=i D^{-1}(z,p)$
satisfies the Schwarz reflection principle, $g^*(z)=g(z^*)$, 
by reintroducing the explicit dependence on 
$\varphi$ and $T$, we can write
\begin{eqnarray}
{\cal F}_j^B\left(\varphi,T\right) &=& 
{1\over2\pi}\int_{-\infty}^{+\infty}
{d p\over 2\pi}\int_{0}^{+\infty}\!\!\!\!\!\! d\omega 
\log\left[i D_j^{-1}(i\omega,p; \varphi,T)\right]
\nonumber\\&&+
\lim_{\epsilon\rightarrow0}
{1\over\pi}\int_{-\infty}^{+\infty}
{d p\over2\pi}\int_0^{+\infty}\!\!\!\!\!\!{d\omega}~
n_{B}(\omega)\left[{\rm arg}\left(
i D_j^{-1}(\omega+i\epsilon,p; \varphi,T)\right)-\pi\right]
\label{formula}
\end{eqnarray}
where $arg(f)$ is the argument $\theta\in[0,2\pi)$ of the complex number
$f\equiv|f|\exp(i\theta)$.

\sect{Free energy density and pressure at order $1/N$}
\indent\noindent

According to the $1/N$ expansion discussed in the previous paragraph, 
${\cal F}$ can be written as the sum of the purely fermionic mean field
term ${\cal F}^F\equiv N{\cal F}_0$, and of the contribution of the bosonic 
fluctuations $\displaystyle\sum_{j=\sigma,\pi}{\cal F}^B_j\equiv {\cal F}_1$
\begin{equation}
{\cal F}(\bar{\varphi}(T),T) = N{\cal F}_0(\bar{\varphi}(T),T) + 
{\cal F}_1(\bar{\varphi}(T),T)
\label{free1n}
\end{equation}
where $\bar{\varphi}(T)$ is the solution of the gap equation (\ref{gap}).
Therefore, in order to get a true $1/N$ expansion of the free 
energy density in
eq.(\ref{free1n}),
we need also to expand $\bar{\varphi}(T)$ in series of $1/N$
\begin{equation}
\bar{\varphi}(T)=\bar{\varphi}_0(T)+{1\over N}\bar{\varphi}_1(T)
+\cdots
\label{expansion}
\end{equation}
By requiring eq.(\ref{gap}) to be satisfied at each order 
in $1/N$,
we obtain that $\bar{\varphi}_0(T)$ is the solution of the 
mean field gap equation
\begin{equation}
\left.{\partial {\cal F}_0(\varphi,T)\over\partial \varphi}\right|
_{\bar{\varphi}_0(T)}=0
\end{equation}
whereas $\bar{\varphi}_1(T)$ is given by
\begin{equation}
\bar{\varphi}_1(T)=-{\displaystyle
\left.{\partial {\cal F}_1(\varphi,T)\over\partial \varphi}
\right|_{\bar{\varphi}_0(T)}}
\cdot\left(
\left.{\partial^2 {\cal F}_0(\varphi,T)\over\partial \varphi^2}
\right|_{\bar{\varphi}_0(T)}\right)^{-1}
\end{equation}
Finally, by inserting eq.(\ref{expansion}) into eq.(\ref{free1n}), we have
\begin{eqnarray}
{\cal F}(\bar{\varphi}(T),T) &=& N{\cal F}_0(\bar{\varphi}_0(T),T) + 
\left.{\partial {\cal F}_0\over\partial \varphi}\right|
_{\bar{\varphi}_0(T)}\!\!\!\!\bar{\varphi}_1(T)+
{\cal F}_1(\bar{\varphi}_0(T),T) + \cdots \nonumber\\[0.4cm]
&=& N{\cal F}_0(\bar{\varphi}_0(T),T) + 
{\cal F}_1(\bar{\varphi}_0(T),T) + \cdots 
\end{eqnarray}

Let us now consider the pressure per fermion species, 
at order $1/N$
\begin{eqnarray}
{\cal P}(T) &=& -{\cal F}(\bar{\varphi}_0(T),T)/N \nonumber\\
&=& {\cal P}_0(\bar{\varphi}_0(T),T) + {1\over N}
{\cal P}_1(\bar{\varphi}_0(T),T) + \cdots
\label{totalpress} 
\end{eqnarray}
According to the previous paragraph the mean field pressure is
given by
\begin{equation}
{\cal P}_0(\bar{\varphi}_0,T) = -{\bar{\varphi}^2_0\over4\pi}
\left(\log{\bar{\varphi}^2_0\over m_0^2} - 1\right)
-\alpha\bar{\varphi}_0 + {2\over\pi\beta}
\int_0^{+\infty}\!\!\!\!\!\!{d p}\log\left(1+e^{\displaystyle 
-\beta\sqrt{p^2+\bar{\varphi}^2_0} }
\right)
\label{meanpress}
\end{equation}
whereas the contribution at  order $1/N$ can be separated into two terms
\begin{eqnarray}
{\cal P}_1\left(\bar{\varphi}_0,T\right) &=&-\sum_{j=\sigma,\pi}
{1\over2\pi}\int_{-\infty}^{+\infty}
{d p\over 2\pi}\int_{0}^{+\infty}\!\!\!\!\!\! d\omega 
\log\left[i D_j^{-1}(i\omega,p; \bar{\varphi}_0,T)\right]
\label{press1n}\\
&&-\sum_{j=\sigma,\pi}
\lim_{\epsilon\rightarrow0}
{1\over\pi}\int_{-\infty}^{+\infty}
{d p\over 2\pi}\int_0^{+\infty}\!\!\!\!\!\!{d\omega}~
n_{B}(\omega)\left[{\rm arg}\left(
i D_j^{-1}(\omega+i\epsilon,p; \bar{\varphi}_0,T)\right)-\pi\right]\nonumber
\end{eqnarray}

In the next section we show the results for the temperature
behaviour of the pressure, expanded as in eq.(\ref{totalpress}),
within the approximation of low temperatures and small current
quark masses.

\sect{Results}
\indent\noindent

Let us first comment on the second term in
eq.(\ref{press1n}), which is the one expected to give the dominant 
bosonic behaviour at low temperatures. Actually such a term contains
the Bose distribution function $n_B(\omega)$, and,
if $iD^{-1}$ were the inverse free bosonic  propagator, 
it  would correspond to the standard pressure 
for a free boson gas \cite{kleva}, and as such it would  contain the
typical temperature behaviour of  a gas of spinless
particles. 

Even in the case of the inverse propagator
having a more complex structure,  the 
major contribution to the pressure 
at low temperatures should come from the integration
over the region containing the poles corresponding to the composite
bosons. In particular, since  
the pion corresponds to the lowest bosonic mass excitation 
because of its pseudogoldstones nature, 
it  characterizes the thermodynamics at 
low temperatures, where the leading behaviours are of the form $\sim
\exp(-m/T)$, for the particles of mass $m$.

Notice that in the present work we are including an explicit
symmetry breaking, avoiding infrared divergencies
related to the low dimensionality of the model. If one
could go continuously to the massless case, one would recover 
the power law behavior in $T$ of the thermodynamical quantities in the
pion sector, evidencing the role of the pions as
Goldstone bosons associated to the spontaneous 
breaking of the chiral symmetry.

In conclusion the pion term in the second sum of eq.(\ref{press1n})
is expected to dominate the total pressure in eq.(\ref{totalpress})
for low temperatures and small values of $\alpha=M/Ng^2$,
which is responsable for the explicit breaking of the chiral symmetry.
In Figs.\ref{alpha01}-\ref{sigmapi} we show that this is indeed the
case, in the present model.

We notice  that the Gross-Neveu model is peculiar 
 since in presence of a bare fermion
mass no bound states are present in the scalar sector (see Appendix $A$).
However this particular feature of the model should not affect its
physical interest to mimic some aspects of QCD 
 at low temperatures, where the contribution of the scalar pole 
is depressed since $m_{\sigma}>\!> m_{\pi}$.

Let us now consider the first term in the r.h.s. of eq.(\ref{press1n}).
Even this term has a temperature dependence,
this time not through the Bose distribution function, but 
 through the Fermi distribution functions appearing in the bosonic
propagator at finite  $T$, obtained by summing over fermions frequencies.

This term can be  separated in a zero temperature part
\begin{equation}
{\cal P}^{0}_{1}\equiv -\sum_{j=\sigma,\pi}
{1\over 2\pi}\int{dp\over 2\pi}\int_{0}^{+\infty}
d\omega ~~{\rm log}\left[i{\cal D}_{0,j}^{-1}
(i\omega,p;\bar{\varphi}_0)\right]
\label{mixzero}
\end{equation}
and a finite temperature part
\begin{equation}
{\cal P}^{\beta}_{1}\equiv  -\sum_{j=\sigma,\pi}
{1\over 2\pi}\int{dp\over 2\pi}\int_{0}^{+\infty}
d\omega ~~{\rm log}\left[1+{{\cal D}_{\beta,j}^{-1}\over
{\cal D}_{0,j}^{-1}} \right]
\label{mixbeta}
\end{equation}
with an analogous separation for the inverse bosonic propagator
(see eq.(\ref{dbeta})
\begin{equation}
{\cal D}^{-1}_j\equiv {\cal D}_{0,j}^{-1}+{\cal D}_{\beta,j}^{-1}
\end{equation} 

${\cal P}^{0}_{1}$ is minus the zero temperature effective
potential of order $1/N$ (evaluated at the minimum $\bar{\varphi}_0$).
It is UV divergent and  needs to be renormalized. The renormalization
procedure for the massless Gross-Neveu model  has been
 studied in literature \cite{root,shon,haym}, although an explicitly
invariant  expression for the  renormalized 
effective potential at this order 
has not been given\footnote{We remind  that since a bare fermion mass
is present, there are no infrared divergencies in the pion 
sector, that otherwise would  be present \cite{root,cole}.}.

However this term (once renormalized) is  subleading, at low
temperatures, with respect to pion pole.
This can be verified by performing a low temperature
expansion, and taking into account that 
${\cal P}_1^0$ depends on temperature 
only through $\bar{\varphi}_0 (T)$, which has the following behaviour 
for low temperatures
\begin{equation}
{\bar{\varphi}}_{0}(T) = {\bar{\varphi}}_{0}(0) +
\delta{\bar{\varphi}}_{0}(T) + \dots
\end{equation}
with
\begin{equation}
\delta{\bar{\varphi}}_{0}(T)\sim \sqrt{T}
e^{\displaystyle{-{{\bar{\varphi}}_{0}(0)\over T}}}
\end{equation}
Thus, as $T$ goes to zero, ${\cal P}_1^0$ is exponentially depressed 
with respect to the pressure term coming from the integration around the 
pion pole, since $\bar{\varphi}_{0}(0)>m_{\pi}$.

Let us now consider ${\cal P}_1^{\beta}$. We have verified numerically
that at  low temperature (and small enough values of $\alpha$) 
also this term is subleading with respect to the pion pole term.
This can be in part understood from the fact that its temperature dependence
comes from the Fermi distribution functions, that, again, 
have an exponentially decreasing behavior at low temperatures, 
characterized by $\bar{\varphi}_{0}(0)$.

In addition both ${\cal P}_1^{0}$ and ${\cal P}_1^{\beta}$ 
 exhibit an imaginary part below
a certain value of the quark condensate. 
 This problem has been already noticed in literature for the
zero temperature effective potential \cite{root,haym},
 and we will not go further in its discussion.
A complete calculation  goes beyond the scope of this paper,
since the problems are specific to the Gross-Neveu
model, whereas the dominance of the pion pole is surely a general feature
of QCD. 

We now discuss the  behaviour of the pure temperature pion 
pressure term discussed above (see second line of eq.(\ref{press1n}))
\begin{equation}
{\cal P}_1^{\pi}\left(\bar{\varphi}_0,T\right)\equiv
-\lim_{\epsilon\rightarrow0}
{1\over\pi}\int_{-\infty}^{+\infty}
{d p\over 2\pi}\int_0^{+\infty}\!\!\!\!\!\!{d\omega}~
n_{B}(\omega)\left[{\rm arg}\left(
i D_{\pi}^{-1}(\omega+i\epsilon,p; \bar{\varphi}_0,T)\right)-\pi\right]
\label{pure}
\end{equation}
and we compare it to the mean field term, eq.(\ref{meanpress}).
The results are shown in Figs.\ref{alpha01}-\ref{alpha001}, 
where we plot various terms of the pressure (divided by the
square temperature) vs $T/m_0$
for different values of the chiral symmetry breaking parameter $\alpha$, 
and of the number $N$ of fermion species.

According to the previous paragraph, and by subtracting the
bag constant ${\cal P}_0(\bar{\varphi}_0(0),0)$, 
the expression for the pressure at order $1/N$  
that we consider is
\begin{equation}
{\cal P}(T)\equiv {\cal P}_0(\bar{\varphi}_0,T)-
{\cal P}_0(\bar{\varphi}_0(0),0)
+{1\over N}{\cal P}_1^{\pi}(\bar{\varphi}_0,T)
\label{totalp}
\end{equation}

Furthermore, to better identify the role of the pion pole,
we have divided the integrations that appear in eq.(\ref{pure})
 into two regions: $(i)$ $p<\omega<\sqrt{p^2+4\bar{\varphi}_0^2}$ 
and $(ii)$ $(\omega^2-p^2)(\omega^2-p^2-4\bar{\varphi}_0^2)>0$.
The region $(i)$ is the one which contains the pion pole, and it is from
here that the pressure takes its major contribution at low temperatures, 
always for small values of $\alpha$. 

In Fig.\ref{alpha01}$(a)$ we show, for $\alpha=0.01m_0$,
the contribution to the pressure
 of the pion term ${\cal P}_{1}^{\pi}$ in eq.(\ref{pure}), 
decomposed as follows:  contribution of the pion
pole (continuous) and that coming from the integration over the region
$(ii)$ (dashed). 
In Fig.\ref{alpha001}$(a)$
 are shown the same quantities of Fig.\ref{alpha01}$(a)$ , 
for $\alpha=0.001m_0$.

As anticipated in the previous paragraphs, we see that at low temperatures
 the pseudoscalar bound state (provided
the current quark mass is small enough) dominates over 
the mean feald pressure, reflecting the fact that the pion is the lowest
mass state by virtue of its nature of pseudogoldstone. 
Its contribution grows more and more as we decrease the bare fermion mass. 
In the massless case 
we would recover a power law behaviour in the limit $T\rightarrow0$,
instead of the exponential one shown in Figs.\ref{alpha01}-\ref{alpha001}
(i.e. the curves of ${\cal P}/T^2$ and the one of ${\cal P}^{\pi}_1/T^2$ 
containing the
contribution of the pion pole would tend to a constant value and not to zero).

In Fig.\ref{alpha01}$(b)$ we plot ${\cal P}/T^2$ (see eq.(\ref{totalp}))
for $N=3$ (short-dashed), $N=10$ (medium-dashed),
and in the mean field case $N=+\infty$ (continuum line). 
In this figure $\alpha=0.01m_0$. 
The analog results for $\alpha=0.001m_0$  are shown
in  Figs.\ref{alpha001}$(b)$. 

Finally, in  Fig.\ref{sigmapi}, we compare ${\cal P}_1^{\pi}$
to the sigma contribution coming from the second term
of eq.(\ref{press1n}) for $\alpha=0.01m_0$ $(a)$ and 
$\alpha=0.001m_0$ $(b)$. Although, as already noticed, the sigma
term is peculiar in this model (since there is no scalar bound state)
we show this figures to make manifest that its contribution is 
negligible at low temperatures, whereas,
 close above the critical temperature $T_c=0.57m_0$, 
the scalar
and pseudoscalar fields have to be practically degenerate, 
since the chiral symmetry
is almost restored (there is no real
restoration due to the explicit symmetry breaking term).

\sect{Conclusions}
\indent\noindent

We have applied a general formula, valid for four-fermion theories, 
to calculate the finite temperature effective potential 
 at order $1/N$ in the massive Gross-Neveu model in $D=1+1$.
We have shown that the pion-like composite
dominates the thermodynamics
at low temperatures, correcting significantly the mean field 
approximation. We have shown in particular that the leading
contribution at low temperatures comes from integration 
close to the pion pole. This feature, which is expected to be 
also general 
to QCD and related models, emerges clearly from our calculations.
\vspace{2cm}

{\sl This work has been carried out within the programm Human Capital 
and Mobility (BBW/OFES 95.0200; CHRXCT 94-0579)}

\newpage
\appendix{
\subsect{Analytic structure of the bosonic propagator}
\indent\noindent

The zero temperature inverse propagator in eq.(\ref{w1})
is given by (we omit the index $j$)
\begin{eqnarray}
D^{-1}_{0}(\omega,p; \varphi)
&=& i\left[ (1+\delta Z) +i\Pi(\omega,p) \right]\\
&=& i\left[ (1+\delta Z) - 2i\lambda
\int{d^2 q\over(2\pi)^2}\displaystyle{q_{\mu}
(q^{\mu}+p^{\mu})\pm \varphi^2\over
\left[(q_{\mu}+p_{\mu})(q^{\mu}+p^{\mu})-\varphi^2\right]
(q_{\mu}q^{\mu} - \varphi^2)}
\right]\nonumber
\end{eqnarray}
where $\pm$ refer to the scalar 
and pseudoscalar self energy $\Pi(\omega,p)$ respectively, and
$p_{\mu}=(\omega,p)$.
By renormalizing in a standard way \cite{gn}, we have
\begin{eqnarray}
&&{2\pi\over \lambda}iD^{-1}_0 (\omega,p;\varphi)=
- \log{\varphi^2\over m_0^2} \nonumber\\
&&\quad+ {(\omega^2-p^2-\epsilon^2_M)\over
\sqrt{(\omega^2-p^2)(\omega^2-p^2-4\varphi^2)}}
\log{\sqrt{(\omega^2-p^2)(\omega^2-p^2-4\varphi^2)}-(\omega^2-p^2)
\over \sqrt{(\omega^2-p^2)(\omega^2-p^2-4\varphi^2)}+(\omega^2-p^2)}
\end{eqnarray}
where $\epsilon^2_{M_{\sigma}} = 4\varphi^2$ and 
$\epsilon^2_{M_{\pi}} = 0$.

At finite temperature the inverse propagator
$D^{-1}$ appearing in eq.(\ref{free}) 
can be calculated by usual methods \cite{land}
\begin{eqnarray}
&&D^{-1} (i\nu_n, p;\varphi,T)=D_0^{-1} (i\nu_n,p;\varphi)+
2i\lambda\int{d q\over 2\pi}
{\displaystyle n_{F}(E_q)\over E_q}\nonumber\\
&&\;-i{\lambda\over2}\left((i\nu_n)^2 - p^2-\epsilon^2_M\right)
\int{d q\over 2\pi}
\left\{
{\displaystyle n_{F}(E_q)\over E_q}\left[{\displaystyle 1\over 
(E_q - i\nu_n)^2 - E_{q+p}^2}+
{\displaystyle 1\over 
(E_q + i\nu_n)^2 - E_{q+p}^2}\right]\right.\nonumber\\
&&\qquad+\left.
{\displaystyle n_{F}(E_{q+p})\over E_{q+p}}\left[{\displaystyle 1\over 
(E_{q+p} - i\nu_n)^2 - E_q^2}+
{\displaystyle 1\over 
(E_{q+p} + i\nu_n)^2 - E_q^2}\right]\right\}
\label{dbeta}
\end{eqnarray}
where  
\begin{equation}
\nu_n=2n\pi /\beta\; ;\quad
n_{F}(E)={1\over e^{\displaystyle \beta E} +1}\; ;\quad
E_q=\sqrt{q^2 + \varphi^2 }
\end{equation}.

Eq.(\ref{dbeta}) can be cast in a more compact form
\begin{eqnarray}
&&{2\pi\over \lambda}iD^{-1} (i\nu_n,p;\varphi,T)=
{2\pi\over \lambda}iD^{-1}_0 (i\nu_n,p;\varphi)
-4\int_0^{+\infty}\!\!\!\!\!\!{dq}
{\displaystyle n_{F}(E_q)\over E_q} \nonumber\\
&&\quad\quad+((i\nu_n)^2-p^2-\epsilon^2_M)\int_0^{+\infty}\!\!\!\!\!\!{dq}
{\displaystyle n_{F}(E_q)\over E_q}\sum_{\xi_1,\xi_2=\pm1}
{\displaystyle 1\over 
(E_q + \xi_1 i\nu_n)^2 - E_{q+\xi_2 p}^2}
\end{eqnarray}

In order to evaluate the second term of eq.(\ref{press1n}) we
take the inverse bosonic propagator as a function of $\omega+i\epsilon$
instead of $i\nu_n$, namely 
$iD^{-1}(\omega+i\epsilon,p;\varphi,T)$. Then, the real and imaginary parts
of $iD^{-1}$ can be evaluated, in the limit $\epsilon\rightarrow0$,
by using
\begin{equation}
\lim_{\epsilon\rightarrow0^{+}}{1\over x+i\epsilon}= PV~{1\over x}
-i\pi\delta(x)
\end{equation}

Defining
$\Delta\equiv{(\omega^2-p^2-4\varphi^2)/(\omega^2-p^2)}$,
we have

\begin{eqnarray}
Re \left[{2\pi\over\lambda}i D^{-1}_{\sigma}\right]
&=&\cases{- \log{\displaystyle\varphi^2\over\displaystyle m_0^2}
-2\sqrt{-\Delta}
\arctan\left({\displaystyle 1\over\displaystyle\sqrt{-\Delta}}\right)
- I^{\beta}+\Delta\; I_1^{\beta}(\omega,p) & $ \Delta < 0 $\cr
&\cr
- \log{\displaystyle\varphi^2\over\displaystyle m_0^2}+\sqrt{\Delta}
\log{\displaystyle\left|\sqrt{\Delta}-1\over\sqrt{\Delta}+1\right|}
- I^{\beta}+\Delta~ I_1^{\beta}(\omega,p)&$\Delta >0 $\cr}
\label{resi}\\[1cm]
Im \left[{2\pi\over\lambda}i D^{-1}_{\sigma}\right]
&=&\cases{0 & $ \Delta < 0 $\cr
&\cr
\displaystyle\pi\sqrt{\Delta}{\displaystyle
\sinh\left({\beta\omega\over 2}\right)\over\displaystyle
\cosh\left({\beta\omega\over 2}\right)+
\cosh\left({\beta p\sqrt{\Delta}\over 2}\right)} &$\Delta >0 $\cr}
\label{imsi}\\[1cm]
Re \left[{2\pi\over\lambda}i D^{-1}_{\pi}\right]
&=&\cases{- \log{\displaystyle\varphi^2\over\displaystyle m_0^2}
+{\displaystyle2\over\displaystyle\sqrt{-\Delta}}
\arctan{\left(\displaystyle 1\over\displaystyle\sqrt{-\Delta}\right)}
- I^{\beta} + I_1^{\beta}(\omega,p) & $ \Delta < 0 $\cr
&\cr
- \log{\displaystyle\varphi^2\over\displaystyle m_0^2}+
{\displaystyle 1\over\displaystyle\sqrt{\Delta}}
\log{\displaystyle\left|\sqrt{\Delta}-1\over\sqrt{\Delta}+1\right|}
 -I^{\beta}+I_1^{\beta}(\omega,p)&$\Delta >0 $\cr}
\label{repi}\\[1cm]
Im \left[{2\pi\over\lambda}i D^{-1}_{\pi}\right]
&=&\cases{0 & $ \Delta < 0 $\cr
&\cr
\displaystyle{\displaystyle\pi\over\displaystyle\sqrt{\Delta}}
{\displaystyle
\sinh\left({\beta\omega\over 2}\right)\over\displaystyle
\cosh\left({\beta\omega\over 2}\right)+
\cosh\left({\beta p\sqrt{\Delta}\over 2}\right)}
 &$\Delta >0 $\cr}\label{impi}
\end{eqnarray}
where we have defined the following integrals
\begin{eqnarray}
I^{\beta}&=&
4\int_0^{+\infty}\!\!\!\!\!\!{dq}
{\displaystyle n_{F}(E_q)\over E_q}\\
I_{1}^{\beta}(\omega,p)&=&
4(\omega^2-p^2)^2\cdot\\
&&\quad\quad\cdot\int_0^{+\infty}\!\!\!\!\!\!{dq}
{\displaystyle n_{F}(E_q)\over E_q}{(\omega^2-p^2)^2-
4\omega^2E^2_q -4p^2q^2\over
\left((\omega^2-p^2)^2-
4\omega^2E^2_q +4p^2q^2\right)^2 - 16p^2q^2(\omega^2-p^2)^2}
\nonumber
\end{eqnarray}

When $\Delta>0$ (and $\omega^2>0$) the last integral 
can be rewritten as follows
\begin{eqnarray}
I_{1}^{\beta}(\omega,p)&=&
{1\over 4\omega p\sqrt{\Delta}}
~PV~\int_0^{+\infty}\!\!\!\!\!\!{dq}
{\displaystyle n_{F}(E_q)\over E_q}\cdot\\
&&\quad\quad\quad\quad\cdot\left((\omega^2-p^2)^2-
4\omega^2E^2_q -4p^2q^2\right)\left(
{1\over q^2-\bar{q}^2_+}-{1\over q^2-\bar{q}^2_-}\right)
\nonumber
\end{eqnarray}
with
$\bar{q}_{\pm}=\left(p\pm \omega\sqrt{\Delta}\right)/2$

Let us now summarize the main features of $iD^{-1}(z,p)$,
with $z=\omega +i\epsilon$,  which 
follow from eqs. (\ref{resi})-(\ref{impi})
\begin {itemize}
\item[a)]
at $T=0$ the inverse propagator $iD^{-1}_0$ has a cut along the
real $z$-axis,
 extending from the continuum threshold
$|\omega|=\sqrt{p^2 + 4\varphi^2}$ to $+\infty$

\item[b)]
for $T\neq0$ a cut is present even for 
$|\omega|<p$. This is due to the fact that at finite temperature  
``scattering'' processes are possible between virtual
scalar/pseudoscalar particles with space like momenta 
$(\omega^2-p^2<0)$ and quarks 
of the medium \cite{kleva,cut}

\item[c)]
in the pseudoscalar sector $iD^{-1}$ has a zero, at least for low temperatures,
for $\omega^2-p^2\sim \alpha m_0 $, corresponding to the pion
bound state \cite{huang}

\item[d)]
no scalar bound states are present for $\alpha\neq0$, both at zero 
\cite{klimenko} and finite temperature~\cite{huang}

\end {itemize}

Finally, we give the expression for the inverse propagator evaluated at 
imaginary energies, which appears in the first term eq.(\ref{press1n})
($\Delta_E\equiv{(\omega^2+p^2+4\varphi^2)/(\omega^2+p^2)}$)
\begin{eqnarray}
{\lambda\over 2\pi}i D^{-1}_{\sigma}(i\omega,p;\varphi,T)
&=&
-\log{\displaystyle\varphi^2\over\displaystyle m_0^2}+\sqrt{\Delta_E}
\log{\displaystyle\left|\sqrt{\Delta_E}-1\over\sqrt{\Delta_E}+1\right|}
- I^{\beta}+\Delta_E~ I_1^{\beta}(i\omega,p)\\[1cm]
{\lambda\over 2\pi}i D^{-1}_{\pi}(i\omega,p;\varphi,T)
&=&- \log{\displaystyle\varphi^2\over\displaystyle m_0^2}+
{1\over\displaystyle\sqrt{\Delta_E}}
\log{\displaystyle\left|\sqrt{\Delta_E}-1\over\sqrt{\Delta_E}+1\right|}
 -I^{\beta}+I_1^{\beta}(i\omega,p)
\end{eqnarray}
for the scalars and pseudoscalars respectively.

\subsect{Sum over bosonic frequencies}
\indent\noindent

The bosonic sums in eq.(\ref{free}) can be transformed into integrals
by means of a standard procedure (for a review of finite temperature
methods see \cite{land}). 
Here we have to pay attention to the fact that, at finite temperature, 
the inverse bosonic propagator $iD^{-1}(z,p)$
(Re$\;z=\omega$) has a cut extending along the real $z$-axis 
 for $-p<\omega<p$.

Let us start by considering the following sum
\begin{equation}
{1\over\beta}\sum_n f(i\nu_n)\; ;\qquad \nu_n=2n\pi/\beta
\end{equation}
where we suppose that 
$f(z)$ has no singularities on the complex plane 
apart from cuts along the real axis. Notice that in our case 
$iD^{-1}(z,p)$, although it has a cut for $-p<\omega<p$, is well
defined in $z=0$, since the discontinuity across the cut
goes to zero as $\omega$ goes to zero.

The first step is then to use
the residue theorem to trade the sum for an integral
over a circuit around the imaginary axis (see Fig.\ref{contour}).
Since the integral along the whole contour 
$\Gamma$ in  Fig.\ref{contour}$(a)$ gives not only the sum
of the residues in $z=i\nu_n$, but also the discontinuity across the
cut, we have to subtract it, namely subtract
the contribution of the contour $C=C_l + C_r$ (horizontal lines).

Thus we can write
\begin{eqnarray}
{1\over\beta}\sum_n f(i\nu_n)&=&
{1\over2\pi i}\left[
\int_{\Gamma} dz f(z)n_{B}(z)
-\int_{C} dz f(z)n_{B}(z)\right] 
\nonumber\\[0.4cm]
&=&\lim_{\delta,\epsilon\rightarrow0}
{1\over2\pi i}\left[
\int_{-i\infty + \delta}^{-i\epsilon + \delta}\!\!\!\!\!\!\!
dz f(z)n_{B}(z)
+\int_{+i\epsilon+ \delta}^{+i\infty+ \delta}\!\!\!\!\!\!\!
dz f(z)n_{B}(z)\right] 
\nonumber\\[0.4cm]
&+&\lim_{\delta,\epsilon\rightarrow0}
{1\over2\pi i}\left[
\int_{+i\infty - \delta}^{+i\epsilon- \delta}\!\!\!\!\!\!\!
dz f(z)n_{B}(z)
+\int_{ -i \epsilon - \delta}^{-i\infty - \delta}\!\!\!\!\!\!\!
dz f(z)n_{B}(z)\right] 
\label{con1}
\end{eqnarray}
where $n_B(z)$ is the Bose distribution function
\begin{equation}
n_{B}(z)={1\over\exp (\beta z)-1}
\end{equation}
which has poles in $z=i\nu_n$, with residue 
equal to $1/\beta$.

The third and fourth integrals of eq.(\ref{con1}) can be transformed
by a change of variable, $z\longrightarrow -z$.
Thus we obtain
\begin{eqnarray}
{1\over\beta}\sum_n f(i\nu_n)&=&
\lim_{\delta,\epsilon\rightarrow0}
{1\over2\pi i}
\int_{-i\infty + \delta}^{-i\epsilon+\delta}\!\!\!\!\!
dz \;n_{B}(z)\left[ f(z) + f(-z)\right]
\nonumber\\[0.4cm]
 &+&
\lim_{\delta,\epsilon\rightarrow0}
{1\over2\pi i}
\int_{+i\epsilon+\delta}^{+i\infty + \delta}\!\!\!\!\!
dz \;n_{B}(z)\left[ f(z) + f(-z)\right]
+{1\over2\pi i}
\int_{-i\infty}^{+i\infty}\!\!\!\!\!dz f(z) 
\label{con2}
\end{eqnarray}
where we have taken into account the fact that no singularities are
encountered in performing the integral of $f(z)$ on the imaginary
axis (last term of eq.(\ref{con2})).

Finally, provided $f(z)\exp(-\beta|z|)$ vanishes sufficiently 
fast at infinity and $f(z)$ has no singularities on the complex plane 
(apart from the already mentioned cuts along the real axis),
we can  rotate the first integrals in eq.(\ref{con2}) 
along the circuit shown in Fig.\ref{contour}$(b)$, and write
\begin{equation}
{1\over\beta}\sum_n f(i\nu_n)=\lim_{\epsilon\rightarrow0}
{1\over\pi i}\int_0^{+\infty}\!\!\!\!\!\!\!d\omega~
n_{B}(\omega)
\left[f(\omega+i\epsilon)-f(\omega-i\epsilon)\right]
+{1\over2\pi}\int_{-\infty}^{+\infty}\!\!\!\!\!\!\! d\omega f(i\omega)
\end{equation}
where we have also supposed $f(z)=f(-z)$, and 
the limit $\delta\rightarrow0$ has been performed.

%end of appendix
}

\newpage

\centerline{\Large\bf Figures captions}

\begin{list}{\bf Fig.\theenumi}{\usecounter{enumi}}

\item{$(a)$ 
Plot of the contribution to the pressure of the
pion term ${\cal P}_{1}^{\pi}$ in eq.(\ref{pure}), 
decomposed as follows:  contribution of the pion
pole (continuous) and that coming from the integration over the region
$(\omega^2-p^2)(\omega^2-p^2-4\bar{\varphi}_0^2)>0$ (dashed). 
These curves are
obtained for $\alpha=0.01m_0$.

$(b)$
Plot of ${\cal P}/T^2$ (see eq.\ref{totalp})
for $N=3$ (short-dashed), $N=10$ (medium-dashed),
and in the mean field case $N=+\infty$ (continuous line). 
In this figure $\alpha=0.01m_0$.
\label{alpha01}}

\item{$(a)$ 
Plot of the contribution to the pressure of the 
pion term ${\cal P}_{1}^{\pi}$ in eq.(\ref{pure}), 
decomposed as follows:  contribution of the pion
pole (continuous) and that coming from the integration over the region
$(\omega^2-p^2)(\omega^2-p^2-4\bar{\varphi}_0^2)>0$ (dashed).
These curves are
obtained for $\alpha=0.001m_0$.

$(b)$
Plot of ${\cal P}/T^2$ (see eq.\ref{totalp})
for $N=3$ (short-dashed), $N=10$ (medium-dashed), 
and in the mean field case $N=+\infty$ (continuous line). 
In this figure $\alpha=0.001m_0$.
\label{alpha001}}

\item{$(a)$ Plot of the contribution to the pressure 
${\cal P}_{1}$ in eq.(\ref{press1n}) of the pion
 (continuous) and of the sigma (dashed).
These curves are
obtained for $\alpha=0.01m_0$.

$(b)$
 Plot of the contribution to the pressure 
${\cal P}_{1}$ in eq.(\ref{press1n}) of the pion
 (continuous) and of the sigma (dashed).
These curves are
obtained for $\alpha=0.001m_0$.
\label{sigmapi}}

\item{$(a)$ 
Contour integration used in eq.(\protect\ref{con1}). $\Gamma$ labels
the whole contour, while $C=C_l + C_r$ is the horizontal 
contour circumscribing the cut across the origin.

$(b)$ Contour integration used in eq.(\protect\ref{con2}). 
$\Gamma'$ labels
the vertical contour along right side of imaginary axis, 
which is transformed in the horizontal 
contours $C_r$ and $C'_r$ 
circumscribing the cuts along the real positive axis.
\label{contour}}

\end{list}
\newpage
\pagestyle{empty}

\begin{figure}[htb]
\setlength{\unitlength}{1mm}
\begin{center}
\begin{picture}(100,80)(-50,-40)
\put(0,-40){\vector(0,1){80}}
\put(-50,0){\vector(1,0){100}}
\put(10,42){\makebox(0,0)[r]{$Im\;\omega$}}  
\put(50,3){\makebox(0,0)[r]{$Re\;\omega$}}  
\multiput(0,-35)(0,5){15}{\circle*{1}}
\put(-10,0){\circle*{1}}
\put(-10,0.5){\line(1,0){20}}
\put(-10,-0.5){\line(1,0){20}}
\put(10,0){\circle*{1}}
\put(20,0){\circle*{1}}
\put(20,0.5){\line(1,0){25}}
\put(20,-0.5){\line(1,0){25}}
\put(-20,0){\circle*{1}}
\put(-20,0.5){\line(-1,0){25}}
\put(-20,-0.5){\line(-1,0){25}}
\put(-1,-2){\makebox(0,0)[r]{$0$}} 
%\put(3,6){\makebox(0,0)[r]{$\delta$}} 
%\put(6,1.8){\makebox(0,0)[r]{$\epsilon$}}
\put(4,-37){\line(0,1){34}}
\put(4,3){\line(0,1){34}}
\put(-4,-37){\line(0,1){34}}
\put(-4,3){\line(0,1){34}}
\put(4,25){\vector(0,1){1}}
\put(-4,-25){\vector(0,-1){1}}
\put(4,0){\oval(18,6)[r]}
\put(8,3){\vector(-1,0){1}}
\put(-4,0){\oval(18,6)[l]}
\put(-8,-3){\vector(1,0){1}}
\put(8,30){\makebox(0,0)[r]{$\Gamma$}} 
\put(12,6){\makebox(0,0)[r]{$C_r$}} 
\put(-6,6){\makebox(0,0)[r]{$C_l$}}
\end{picture}
\end{center}
\end{figure}
\centerline{(a)}

\begin{figure}[htb]
\setlength{\unitlength}{1mm}
\begin{center}
\begin{picture}(100,80)(-50,-40)
\put(0,-40){\vector(0,1){80}}
\put(-50,0){\vector(1,0){100}}
\put(10,42){\makebox(0,0)[r]{$Im\;\omega$}}  
\put(55,3){\makebox(0,0)[r]{$Re\;\omega$}}  
\multiput(0,-35)(0,5){15}{\circle*{1}}
\put(-10,0){\circle*{1}}
\put(-10,0.5){\line(1,0){20}}
\put(-10,-0.5){\line(1,0){20}}
\put(10,0){\circle*{1}}
\put(20,0){\circle*{1}}
\put(20,0.5){\line(1,0){25}}
\put(20,-0.5){\line(1,0){25}}
\put(-20,0){\circle*{1}}
\put(-20,0.5){\line(-1,0){25}}
\put(-20,-0.5){\line(-1,0){25}}
\put(-1,-2){\makebox(0,0)[r]{$0$}} 
%\put(3,6){\makebox(0,0)[r]{$\delta$}} 
%\put(6,1.8){\makebox(0,0)[r]{$\epsilon$}}
\put(4,-37){\line(0,1){34}}
\put(4,3){\line(0,1){34}}
\put(4,25){\vector(0,1){1}}
\put(4,0){\oval(18,6)[r]}
\put(8,3){\vector(1,0){1}}
\put(45,0){\oval(56,6)[l]}
\put(40,3){\vector(1,0){1}}
\put(4,3){\oval(82,68)[tr]}
\put(4,-3){\oval(82,68)[br]}
\put(9,30){\makebox(0,0)[r]{$\Gamma'$}} 
\put(14,6){\makebox(0,0)[r]{$-C_r$}} 
\put(35,6){\makebox(0,0)[r]{$C_r'$}}
\end{picture}
\end{center}
\end{figure}
\centerline{(b)}\vspace{1cm}
\centerline{\Large \bf Fig.\protect\ref{contour}}

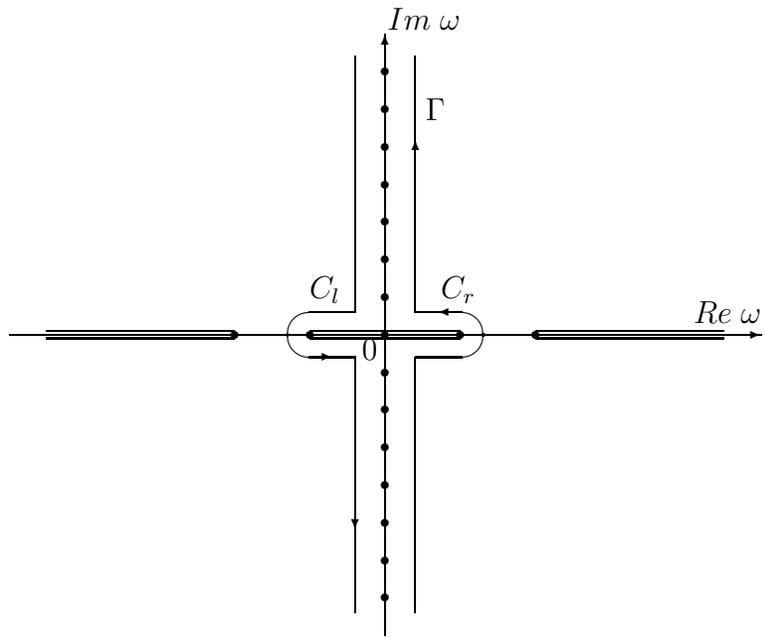
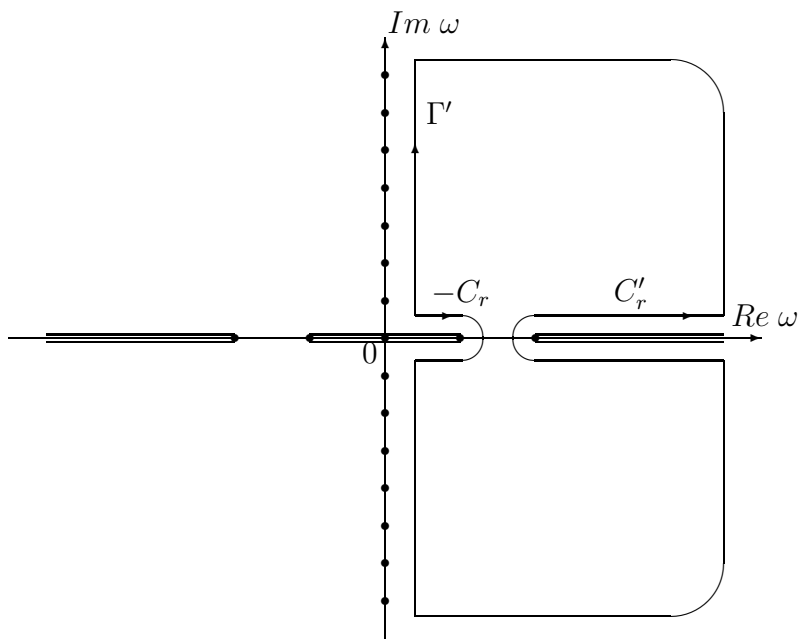
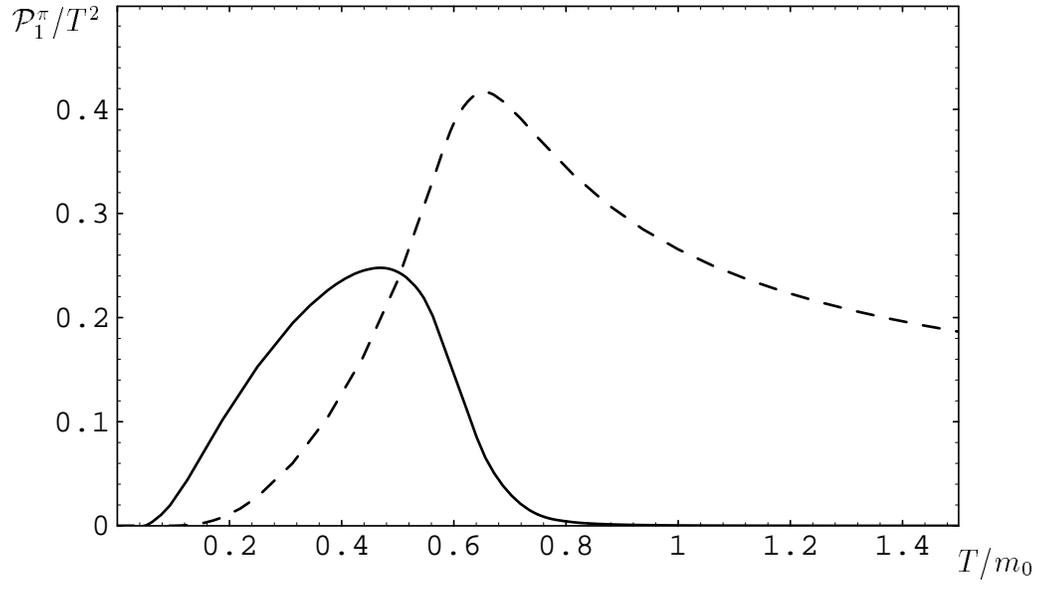
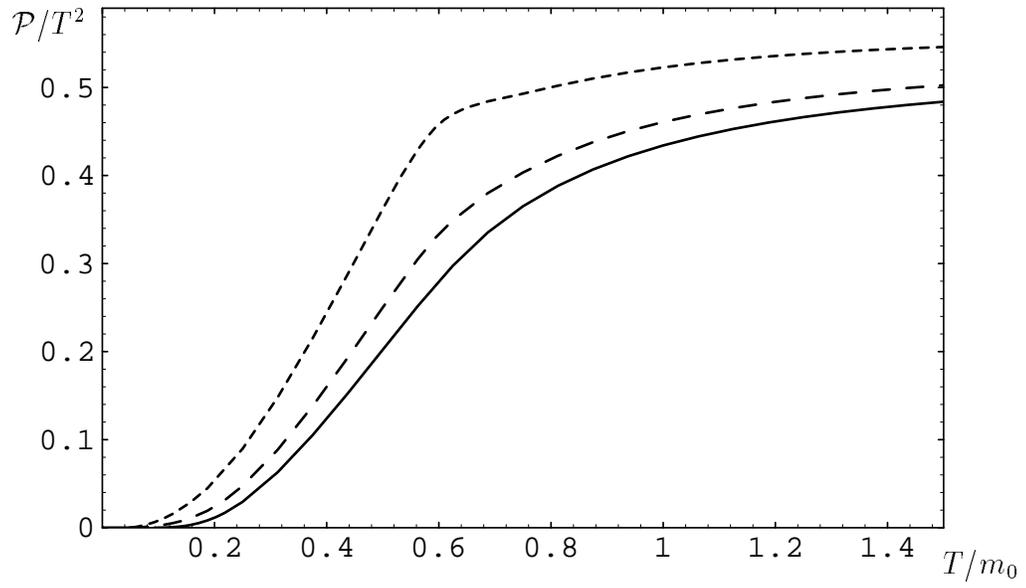
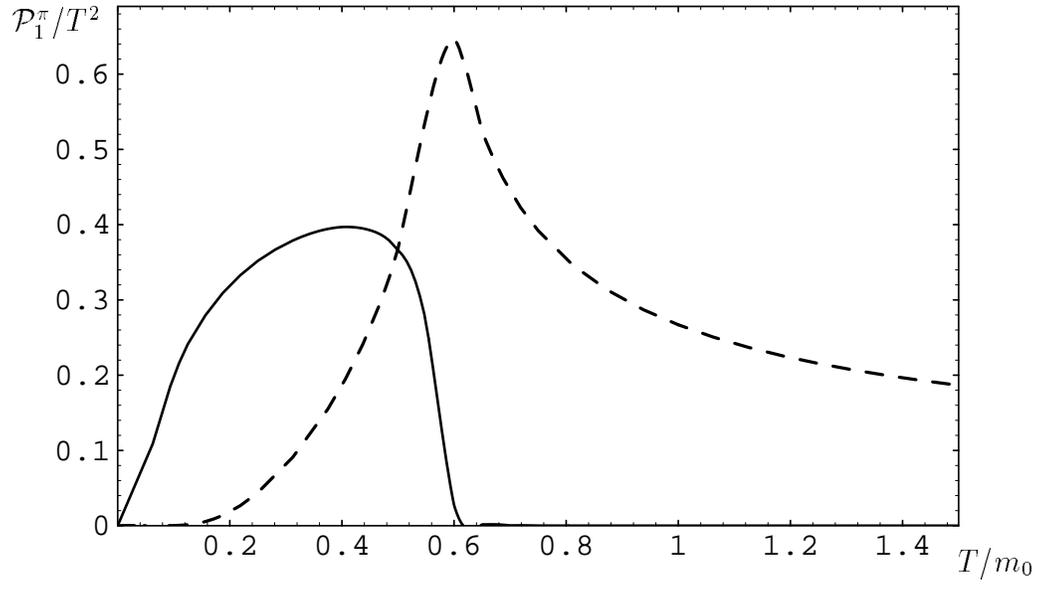
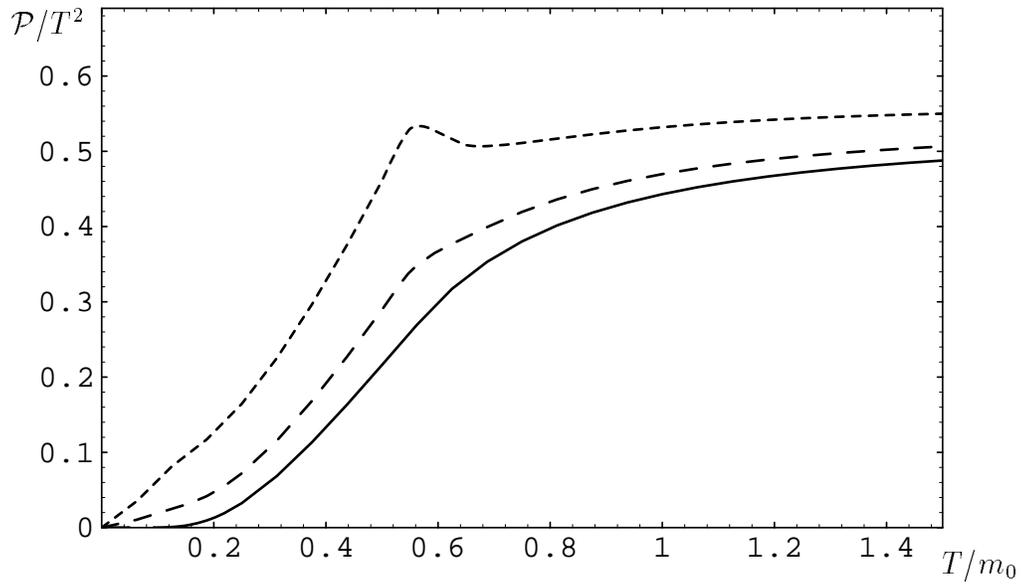
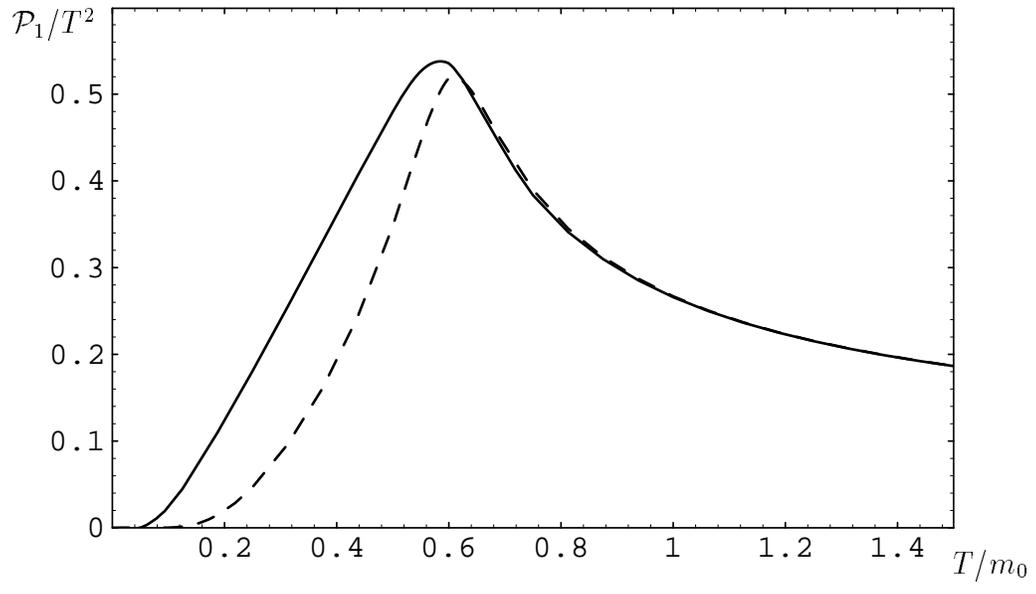
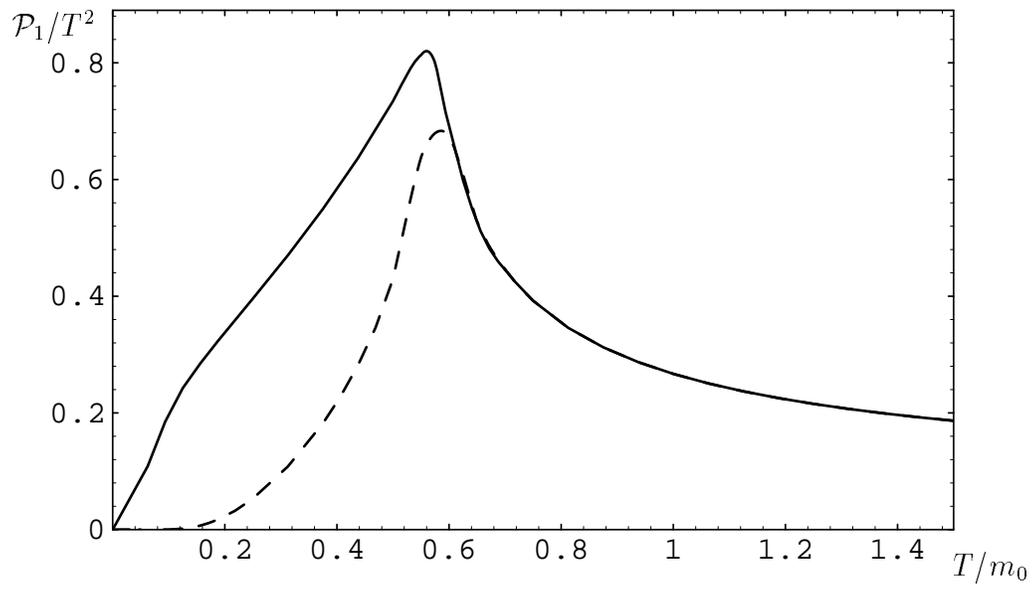
\end{document}